\documentclass{lpm2ws}

\begin{document}

\title{Integral characteristics
of bremsstrahlung and pair photoproduction in a medium}

\author{V. N. Baier and V. M. Katkov}

\address{Budker Institute of Nuclear Physics, Novosibirsk, 630090,
Russia \\E-mail: baier@inp.nsk.su}


\maketitle

\abstracts{The bremsstrahlung of an electron and $e^{-}e^{+}$-pair
creation by a photon in a medium is considered in high-energy
region, where influence of the multiple scattering on the
processes (the Landau-Pomeranchuk-Migdal (LPM) effect) becomes
essential. The integral characteristics: the radiation length and
the total probability of radiation and pair photoproduction are
analyzed under influence of the LPM effect.}

\section{Introduction}

When a charged particle is moving in a medium it scatters on atoms.
With probability $\sim \alpha$ this scattering is accompanied by a radiation.
At high energy the radiation process occurs over a rather long distance,
known as the {\em formation length} $l_c$:
\begin{equation}
l_c=\frac{l_0}{1+\gamma^2 \vartheta_c^2},\quad
l_0=\frac{2\varepsilon \varepsilon'}{m^2\omega},
\label{1.1a}
\end{equation}
where $\omega$ is the energy of emitted photon,
$\varepsilon (m)$ is the energy
(the mass) of a particle, $\gamma=\varepsilon/m$ is the Lorenz factor,
$\varepsilon'=\varepsilon-\omega$,
$\vartheta_c$ is the characteristic angle of photon emission,
the system $\hbar=c=1$ is used.

Landau and Pomeranchuk were the first who showed that if the formation
length of bremsstrahlung becomes comparable to the distance over which
the multiple scattering becomes important (when the mean angle
of multiple scattering is of the order of the characteristic angle of
photon emission $\sim 1/\gamma$), the bremsstrahlung will be
suppressed \cite{1}. Migdal \cite{M1} developed the quantitative
theory of this phenomenon.

New activity with the theory of the LPM effect
(see \cite{BD}, \cite{BDM}, \cite{Z})
is connected with a very
successful series of experiments performed
at SLAC recently (see \cite{E1}, \cite{E2}).
In these experiments the cross section
of the bremsstrahlung of soft photons with energy from 200~keV to
500~MeV from electrons with energy 8~GeV and 25~GeV is measured
with an accuracy of the order of a few percent. Both LPM and dielectric
suppression are observed and investigated. These experiments were the
challenge for the theory since in all the mentioned papers calculations
are performed to logarithmic accuracy which is not enough for description
of the new experiment. The contribution of the Coulomb corrections (at least
for heavy elements) is larger than experimental errors and these corrections
should be taken into account.

We developed the new approach to the theory of the
Landau-Pomeranchuk-Migdal (LPM) effect \cite{L1} basing on the
quasiclassical operator approach \cite{BKS}. In this paper
the cross section of the bremsstrahlung process in the photon
energies region where the influence of the LPM is very strong was
calculated with a term $\propto 1/L$ , where $L$ is
characteristic logarithm of the problem, and with the Coulomb
corrections taken into account. In the photon energy region,
where the LPM effect is "turned off", the obtained cross section
gives the exact Bethe-Maximon cross section (within power
accuracy) with the Coulomb corrections. This important feature
was absent in the previous calculations. Some important features
of the LPM effect were considered also in \cite{L2}, \cite{L3},
\cite{L4}, \cite{L5}.

The crossing process for the bremsstrahlung is the pair creation
by a photon. The created particles undergo here the multiple
scattering. It should be emphasized that for the bremsstrahlung
the formation length (\ref{1.1a}) increases strongly if $\omega
\ll \varepsilon$. Just because of this the LPM effect was
investigated at SLAC at a relatively low energy. For the pair
creation by a photon with energy $\omega$ the formation length
$\displaystyle{l_{p}
=\frac{2\varepsilon(\omega-\varepsilon)}{m^2\omega}}$ attains
maximum at $\varepsilon=\omega/2$ and this maximum is
$l_{p,max}=(\omega/2m)\lambda_c$. Because of this even for heavy
elements the effect of multiple scattering becomes noticeable at
photon energies $\omega \geq 10$~TeV. Starting from these
energies one has to take into account the influence of a medium
on the pair creation and on the bremsstrahlung hard part of the
spectrum in electromagnetic showers being created by the cosmic
ray particles of the ultrahigh energies. These effects can be
quite significant in the electromagnetic calorimeters operating
in the detectors on the colliders in TeV range.

In the present paper the radiation length is calculated under
influence of the LPM effect. The total probability of photon
radiation and the integral probability of the pair creation are
considered also.

\section{Influence of the multiple scattering
on the bremsstrahlung}

\subsection{Bremsstrahlung spectrum at high energy}

The spectral radiation intensity obtained in \cite{L1} (see Eq.(2.39))
has the form
\begin{equation}
dI=\omega dW=\frac{\alpha m^2 xdx}{2\pi (1-x)}
{\rm Im}~\left[\Phi(\nu)-\frac{1}{2L_c}F(\nu)
\right],\quad x=\frac{\omega}{\varepsilon},
\label{3.1}
\end{equation}
where
\begin{eqnarray}
&&\displaystyle{\Phi(\nu)=\int_{0}^{\infty} dz e^{-it}\left[r_1
\left(\frac{1}{\sinh z}-\frac{1}{z}\right)-i\nu r_2
\left( \frac{1}{\sinh^2z}- \frac{1}{z^2}\right) \right]}
\nonumber \\
&& = r_1\left(\ln p-\psi\left(p+\frac{1}{2}\right) \right)
+r_2\left(\psi (p) -\ln p+\frac{1}{2p}\right),
\nonumber \\
&&F(\nu)= \int_{0}^{\infty}\frac{dz e^{-it}}{\sinh^2z}
\left[r_1f_1(z)-2ir_2f_2(z) \right],
\nonumber \\
&& f_1(z)=\left(\ln \varrho_c^2+\ln \frac{\nu}{i}
-\ln \sinh z-C\right)g(z) - 2\cosh z G(z),
\nonumber \\
&& f_2(z) = \frac{\nu}{\sinh z}
\left(f_1(z)-\frac{g(z)}{2} \right),\quad g(z)=z\cosh z - \sinh z,
\nonumber \\
&& G(z)=\int_{0}^{z}(1-y\coth y)dy
\nonumber \\
&&\displaystyle{=z-\frac{z^2}{2}-\frac{\pi^2}{12}-
z\ln \left(1-e^{-2z} \right)
+\frac{1}{2}{\rm Li}_2 \left(e^{-2z} \right)},
\nonumber\\
&& t=\frac{z}{\nu},\quad r_1=x^2,\quad r_2=1+(1-x)^2,\quad
t=t_1+t_2,~z=\nu t.
\label{3.2}
\end{eqnarray}
here $\alpha=1/137,~z=\nu t,~p=i/(2\nu),~\psi(x)$ is the
logarithmic derivative of the gamma function, ${\rm Li}_2 \left(x
\right)$ is the Euler dilogarithm. Use of found form of $\Phi$
and the last representation of function $G(z)$ simplifies the
numerical calculation. The term with $\Phi(\nu)$ in (\ref{3.1})
describes the intensity in logarithmic approximation, the term
with $F(\nu)$ is the first correction. The parameters in these
formulas are
\begin{eqnarray}
&& \nu^2=i\nu_0^2,\quad \nu_0^2=|\nu|^2 \simeq \nu_1^2\left(1+
\frac{\ln \nu_1}{L_1}\vartheta(\nu_1-1) \right),\quad
\nu_1^2=\frac{\varepsilon}{\varepsilon_e}\frac{1-x}{x},
\nonumber \\
&&\varepsilon_e=m\left(8\pi Z^2 \alpha^2 n_a \lambda_c^3 L_1
\right)^{-1},~ L_c \simeq L_1 \left(1+ \frac{\ln
\nu_1}{L_1}\vartheta(\nu_1-1) \right),~ L_1=\ln
\frac{a_{s2}^2}{\lambda_c^2},
\nonumber \\
&&\frac{a_{s2}}{\lambda_c}=183Z^{-1/3}e^{-f},\quad
f=f(Z\alpha)=(Z\alpha)^2\sum_{k=1}^{\infty}\frac{1}{k(k^2+(Z\alpha)^2)},
\label{3.3}
\end{eqnarray}
here $Z$ is the charge of the nucleus, $n_a$ is the number
density of atoms in the medium, $\lambda_c=1/m$ is the electron
Compton wavelength. The LPM effect manifests itself when
\begin{equation}
\nu_1(x_c)=1,\quad x_c=\frac{\varepsilon}{\varepsilon_e+\varepsilon} .
\label{3.4}
\end{equation}

In the case $\varepsilon \ll \varepsilon_e$ in the hard part of spectrum
($1 \geq x \gg x_c$) the parameter $\nu_1^2 \simeq x_c/x \ll 1$ and the
contribution into the integral (\ref{3.2}) give the region $z \ll 1$.
\begin{equation}
\displaystyle{{\rm Im}~\Phi(\nu) \simeq r_1\frac{\nu_1^2}{6}
+r_2\frac{\nu_1^2}{3},\quad
 -{\rm Im}~F(\nu)= -\frac{1}{9}(r_2-r_1)\nu_1^2 (1 + O(\nu_1^4))}.
\label{3.5}
\end{equation}
Substituting into (\ref{3.1}) we have
\begin{equation}
\frac{dI}{dx}=\frac{2Z^2\alpha^3n_a\varepsilon}{3m^2}\Bigg[r_1
\left(L_1-\frac{1}{3} \right)
+2r_2 \left(L_1+\frac{1}{6} \right)\Bigg]
\label{3.6}
\end{equation}
This is the Bethe-Maximon intensity spectrum (with the Coulomb
corrections) in case of complete screening (if one neglects the
contribution of atomic electrons) written down within power
accuracy (omitted terms are of the order of powers of $1/\gamma$),
see e.g. Eq.(18.30) in \cite{BKF}. So, to obtain it in the limit
considered one has to take into account the both terms in
brackets in (\ref{3.1}).

At very strong multiple scattering $\nu_0 \gg 1$ or
$\varepsilon \gg \varepsilon_e$ one can omit $e^{-it}$
in the integrand of function $F_(\nu)$
(\ref{3.2}).
Integrating over $z$ we obtain
\begin{equation}
\displaystyle{-{\rm Im}~F(\nu)=\frac{\pi}{4}(r_1-r_2)+\frac{\nu_0}{\sqrt{2}}
\left(\ln 2-C+\frac{\pi}{4} \right)r_2},
\label{2.25}
\end{equation}
where we take into account the next terms of the decomposition in the
term $\propto r_2$.
Under the same conditions ($\nu_0 \gg 1$) the function ${\rm Im}~\Phi(\nu)$
is
\begin{equation}
\displaystyle{{\rm Im}~\Phi(\nu)=\frac{\pi}{4}(r_1-r_2)+
\frac{\nu_0}{\sqrt{2}}r_2}.
\label{2.26}
\end{equation}
Thus, at $\nu_0 \gg 1$ the relative contribution of the first correction
$\displaystyle{\frac{dW^1}{d\omega}}$ is defined by
\begin{equation}
\displaystyle{r=\frac{dW^1}{dW^c}=\frac{1}{2L_c}
\left(\ln 2-C+\frac{\pi}{4} \right) \simeq \frac{0.451}{L_c}}.
\label{2.27}
\end{equation}

In the case $\varepsilon \geq \varepsilon_e$ the intensity
spectrum differs from the Bethe-Maximon  one at $x \sim 1$ also.
When $\varepsilon \gg \varepsilon_e$ we find in the interval not
very close to the end of the spectrum ($x=1$):
\begin{eqnarray}
&&\frac{dI}{dx} \simeq \frac{2\sqrt{2}Z^2\alpha^3n_a\varepsilon}{m^2}
\sqrt{\frac{\varepsilon_e x}{\varepsilon (1-x)}}
\left(1+\frac{1}{4L_1}\ln \frac{\varepsilon (1-x)}{\varepsilon_e x} \right)
\Bigg[x^2\nonumber \\
&& +2(1-x)\left(1-\frac{\pi}{2\sqrt{2}}
\sqrt{\frac{\varepsilon_e x}{\varepsilon (1-x)}}\right) \Bigg],\quad
\varepsilon(1-x) \gg \varepsilon_ex.
\label{3.7}
\end{eqnarray}

\subsection{Integral characteristics of bremsstrahlung}

Now we turn to the integral characteristics of radiation.
The total intensity of radiation in the logarithmic approximation
can be presented as (see (\ref{3.1}))
\begin{eqnarray}
&&\frac{I}{\varepsilon}L_{rad}^0=2\frac{\varepsilon_e}{\varepsilon}
{\rm Im}\Bigg[\int_{0}^{1}\frac{dx}{g}\sqrt{\frac{x}{1-x}}(2(1-x)+x^2)
\nonumber \\
&&+ \int_{0}^{1}\frac{x^3dx}{1-x}
\left(\psi(p+1)-\psi\left(p+\frac{1}{2}\right) \right)
+2\int_{0}^{1}xdx\left(\psi\left(p+1\right)
-\ln p\right)\Bigg],
\label{3.7a}
\end{eqnarray}
where
\[
p=\frac{g\eta}{2},\quad \eta=\sqrt{\frac{x}{1-x}},\quad
g=\exp \left(i\frac{\pi}{4}\right)
\sqrt{\frac{L_1}{L_c}\frac{\varepsilon_e}{\epsilon}},
\]
$L_{rad}^0$ is the radiation length in the logarithmic
approximation. The relative energy losses of electron per unit
time in terms of the Bethe-Maximon radiation length $L_{rad}^0$:
$\displaystyle{\frac{I}{\varepsilon}L_{rad}^0}$ in gold is given
in Fig.1 (curve 1), it reduces by 10\% (15\% and 25\%) at
$\varepsilon \simeq 700$~GeV ($\varepsilon \simeq 1.4$~TeV and
$\varepsilon \simeq 3.8$~TeV) respectively, and it cuts in half
at $\omega \simeq 26$~TeV. This increase of effective radiation
length can be important in electromagnetic calorimeters operating
in detectors on colliders in TeV range. The contribution of the
correction terms $r$ (see (\ref{2.27})) is $r \simeq 0.451/L_c$.

In Eqs.(\ref{3.6}) and (\ref{3.7}) we can use the main terms of
decomposition only.

\begin{figure}[t]
\vspace*{-2.3cm}
\epsfxsize=20pc 
{\centering \epsfbox{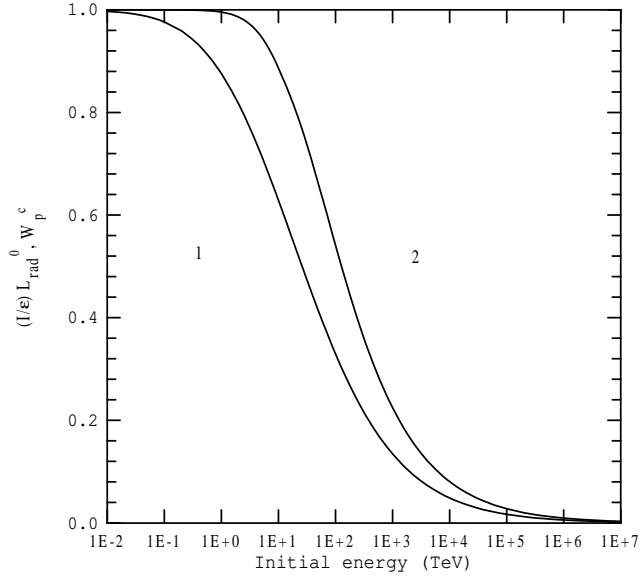} }
\vskip-2.cm \caption{The relative energy losses of electron per
unit time in terms of the Bethe-Maximon radiation length
$L_{rad}^0$: $\displaystyle{\frac{I}{\varepsilon}L_{rad}^0}$ in
gold Eq.(\ref{3.7a}) vs the initial energy of electron (curve 1)
and the total pair creation probability per unit time $W_p^c$ (see
Eq.(\ref{2.23b}))in terms of the Bethe-Maximon total probability
of pair creation $W_{p0}^{BH}$ in gold vs
the initial energy of photon (curve 2). \label{fig:radish}}
\end{figure}

The main term in (\ref{3.6}) gives after the integration over $x$
the standard expression for the radiation length $L_{rad}$
without influence of multiple scattering.
\begin{eqnarray}
&&\frac{I}{\varepsilon}=\frac{\alpha m^2}{4\pi \varepsilon_e}
\left(1+\frac{1}{9L_1}-\frac{4\pi}{15}\frac{\varepsilon}{\varepsilon_e}
\right) \simeq L_{rad}^{-1}\left(1
-\frac{4\pi}{15}\frac{\varepsilon}{\varepsilon_e}\right),
\nonumber \\
&& \frac{1}{L_{rad}}= \frac{2Z^2\alpha^3n_a L_1}{m^2}
\left(1+\frac{1}{9L_1} \right)=
\frac{1}{L_{rad}^0}\left(1+\frac{1}{9L_1} \right)
\label{3.8}
\end{eqnarray}

The integration over $x$  of the main term in (\ref{3.7}) gives
(terms $\propto \sqrt{\varepsilon_e/\varepsilon}$ in the square brackets
are neglected)
\begin{equation}
I_0 \simeq \frac{9\pi Z^2 \alpha^3 n_a \sqrt{\varepsilon
\varepsilon_e}} {4\sqrt{2}m^2} L_1 \left[1+\frac{1}{4L_1}
\left(\ln \frac{\varepsilon}{\varepsilon_e}-\frac{46}{27}
\right)+r_0 \right],
\label{3.9}
\end{equation}
where $ r_0=\left(\ln 2 -C +\pi/4 \right)/2L_1$. The corrections
(without terms $\propto 1/L_1$) to (\ref{3.9}) are calculated in
Appendix B of \cite {L5}(see Eq.(B.11)). The complete result is
\begin{equation}
\frac{I}{\varepsilon L_{rad}} \simeq \frac{5}{2}
\sqrt{\frac{\varepsilon_e}{\varepsilon}}
\left[1-2.37\sqrt{\frac{\varepsilon_e}{\varepsilon}} -4.57
\frac{\varepsilon_e}{\varepsilon} +\frac{1}{4L_1} \left(\ln
\frac{\varepsilon}{\varepsilon_e} -0.3455 \right)\right]
\label{3.10}
\end{equation}
Although the coefficients in the last expression are rather large at
two first terms of the decomposition over
$\sqrt{\varepsilon_e/\varepsilon}$ this formula has the accuracy of the order
of 10\% at $\varepsilon \sim 10 \varepsilon_e$.

The integral probability of radiation in terms of the
Bethe-Maximon radiation length can be obtained from
Eq.(\ref{3.7a}) dividing the integrand in all integrals by $x$. It
is given in Fig.2. It should be mentioned that the standard
Bethe-Maximon integral probability doesn't exist at all (the
integral over $\omega$ has the logarithmic divergence at $\omega
\rightarrow 0$). Due to the LPM effect the soft part of the
spectrum is damped and integral over $\omega$ exists.

The integral probability of radiation for $\varepsilon \ll
\varepsilon_e$ was calculated in \cite{L3}:
\begin{eqnarray}
&&W=\frac{4}{3L_{rad}^0}\left(\ln
\frac{\varepsilon_e}{\varepsilon}+C_2 \right), \nonumber \\
&& C_2=2C-\frac{5}{8}+12 \int_{0}^{\infty}\ln z
\left(\frac{1}{z^3}-\frac{\cosh z}{\sinh^3 z} \right)dz \simeq
1.96
\label{3.11}
\end{eqnarray}

In the case $\varepsilon \gg \varepsilon_e$
we can calculate the integral probability of radiation starting with
Eq.(\ref{3.7}). Conserving the main term, dividing it by $x\varepsilon$
and integrating over $x$ we find
\begin{equation}
W_0=\frac{11\pi Z^2 \alpha^3 n_a}{2 \sqrt{2} m^2}
\sqrt{\frac{\varepsilon_e}{\varepsilon}} L_1 \left[1+
\frac{1}{4L_1} \left(\ln \frac{\varepsilon}{\varepsilon_e}
+\frac{8}{11} \right)+r_0 \right]
\label{3.12}
\end{equation}

The correction terms to Eq.(\ref{3.11}) are calculated in
Appendix B of \cite{L5}(see Eq.(B.13)). Substituting them we have
\begin{equation}
W=\frac{11\pi Z^2 \alpha^3 n_a}{2\sqrt{2} m^2}
\sqrt{\frac{\varepsilon_e}{\varepsilon}}L_1
\left[1-1.23\sqrt{\frac{\varepsilon_e}{\varepsilon}}
+1.65 \frac{\varepsilon_e}{\varepsilon}
+\frac{1}{4L_1}
\left(\ln \frac{\varepsilon}{\varepsilon_e}
+2.53 \right)\right].
\label{3.13}
\end{equation}

Ratio of the main terms of Eqs.(\ref{3.10}) and (\ref{3.13})
gives the mean energy of radiated photon
\begin{equation}
\bar{\omega}=\frac{9}{22}\varepsilon \simeq 0.409 \varepsilon.
\label{3.14}
\end{equation}

\section{Influence of multiple scattering on pair creation process}

The probability of the pair creation by a photon can be obtained
from the probability of the bremsstrahlung with help of the
substitution law:
\begin{equation}
\omega^2d\omega \rightarrow \varepsilon^2 d\varepsilon,\quad
\omega \rightarrow -\omega,\quad \varepsilon \rightarrow
-\varepsilon, \label{2.1}
\end{equation}
where $\omega$ is the initial photon energy, $\varepsilon$ is the
energy of the created electron. Making this substitution in
Eq.(\ref{3.1}) we obtain the spectral distribution of the pair
creation probability (over the energy of the electron)
\begin{eqnarray}
&&\displaystyle{\frac{dW_p^c}{d\varepsilon}= \frac{\alpha
m^2}{2\pi \varepsilon \varepsilon'} {\rm Im}~\left[\Phi_p
(\nu)-\frac{1}{L_c}F_p(\nu)\right]},
\nonumber \\
&&\displaystyle{\Phi_p(\nu)=\nu\int_{0}^{\infty} dt
e^{-it}\left[s_1 \left(\frac{1}{\sinh z}-\frac{1}{z}\right)-i\nu
s_2 \left( \frac{1}{\sinh^2z}- \frac{1}{z^2}\right) \right]}
\nonumber \\
&&=s_1\left(\ln p-\psi\left(p+\frac{1}{2}\right) \right)
+s_2\left(\psi (p) -\ln p+\frac{1}{2p}\right), \nonumber \\
&& F_p(\nu)= \int_{0}^{\infty}\frac{dz e^{-it}}{\sinh^2z}
\left[s_1f_1(z)-2is_2f_2(z) \right], \nonumber \\
&& s_1=1,\quad s_2=\frac{\varepsilon^2+\varepsilon'^2}
{\omega^2},\quad\varepsilon'=\omega-\varepsilon.
 \label{2.10}
\end{eqnarray}
All entering functions are defined in (\ref{3.2}).

The total probability of pair creation in the logarithmic
approximation can be presented as (see (\ref{2.10}))
\begin{eqnarray}
&&\frac{W_p^c}{W_{p0}^{BH}}=\frac{9}{14}\frac{\omega_e}{\omega}
{\rm Im}\int_{0}^{1}\frac{dy}{y(1-y)}\Bigg[ \left(\ln
p-\psi\left(p+\frac{1}{2}\right) \right)
\nonumber \\
&&+\left(1-2y+2y^2 \right)\left(\psi\left(p\right) -\ln
p+\frac{1}{2p}\right)\Bigg], \quad p=\frac{bs}{4},\label{2.23b}
\end{eqnarray}

\begin{figure}[t]
\vspace*{-2.3cm}
\epsfxsize=20pc 
{\centering \epsfbox{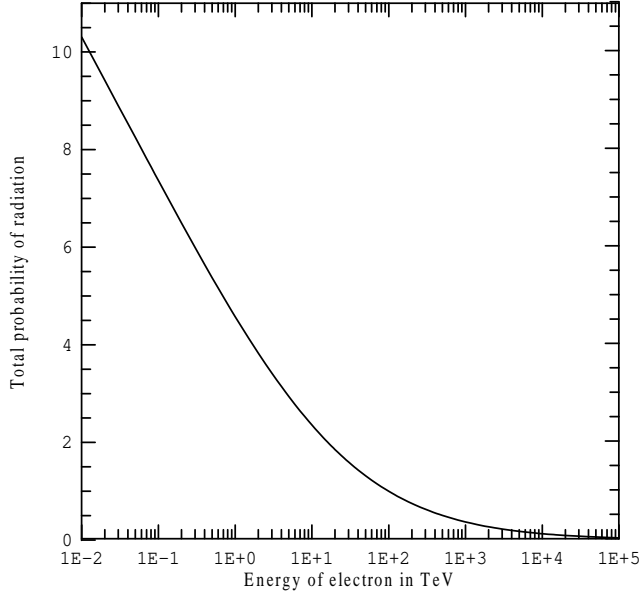} }
\vskip-2.cm \caption{The total probability of photon emission
$W_0$ in terms of the Bethe-Maximon radiation length $L_{rad}^0$
in gold vs the initial energy electron . \label{fig:2}}
\end{figure}

\hspace*{-0.6cm}where
\[
 s=\frac{1}{\sqrt{y(1-y)}},\quad b=\exp
\left(i\frac{\pi}{4}\right)
\sqrt{\frac{L_1}{L_c}\frac{\omega_e}{\omega}},\quad
\omega_e=m\left(2\pi Z^2 \alpha^2 n_a \lambda_c^3 L_1
\right)^{-1},
\]
here $W_{p0}^{BH}$ is the Bethe-Maximon probability of pair
photoproduction in the logarithmic approximation. Note that
$\omega_e$ is four times larger than $\varepsilon_e$, in gold
$\omega_e=10.5~$TeV. This is just the value of photon energy
starting with the LPM effect becomes essential for the pair
creation process in heavy elements. The total probability of pair
creation $W_p^c$ in gold is given in Fig.1 (curve 2),it reduced
by 10\% at $\omega \simeq 9$~TeV and it cuts in half at $\omega
\simeq 130$~TeV.

\section{Conclusion}

In this paper we considered the influence of multiple scattering on
the bremsstrahlung process at any energy including
the high-energy region ($\varepsilon \geq
\varepsilon_e$), where all the spectrum
of radiation is distorted.
In this region the total intensity of radiation diminishes and
respectively the radiation length increases. The cross section of
$e^-e^+$ pair creation by a photon changes essentially if the photon energy
$\omega \geq \omega_e=4\varepsilon_e$, see Eq.(\ref{3.3}).

If we restrict to the main terms of the decomposition Eq.(\ref{3.10})
in asymptotic region $\varepsilon \gg \varepsilon_e$,
then the intensity of radiation and the corresponding radiation length
can be written as
\begin{equation}
I \simeq \frac{9}{16}\sqrt{\frac{\pi}{2}}Z\alpha^2
\left(\varepsilon n_a \ln \left(9\pi Z^2\alpha^2\varepsilon n_a
a_{s2}^4 \right) \right)^{1/2},\quad
L_{rad}=\frac{\varepsilon}{I(\varepsilon)}.
\label{4.1}
\end{equation}
The integral cross section of radiation follows from the
 integral probability of radiation (\ref{3.13})
\begin{equation}
\sigma = \frac{W}{n_a} \simeq \frac{11}{8}\sqrt{\frac{\pi}{2}}
\frac{Z\alpha^2}{\sqrt{\varepsilon n_a}}
\left(\ln \left(100\pi Z^2\alpha^2\varepsilon n_a
a_{s2}^4 \right) \right)^{1/2}.
\label{4.2}
\end{equation}
We have from for the total probability
of pair creation by a photon at $\omega \gg \omega_e$ and the
corresponding cross section
\begin{equation}
W_p \simeq \frac{3}{4}\sqrt{\frac{\pi}{2}}
Z\alpha^2
\left(\frac{n_a}{\omega}\ln \left(2\pi Z^2\alpha^2\omega n_a
a_{s2}^4 \right) \right)^{1/2},\quad \sigma_p=\frac{W_p}{n_a}
\label{4.3}
\end{equation}
The Eqs.(\ref{4.1})-(\ref{4.3}) don't depend on the electron mass
and the cross sections of bremsstrahlung and pair creation
diminish with energy and density $n_a$ growth.

In this paper we considered the case of an infinitely thick target
where the formation length is much shorter than the thickness of
a target. Because of this we neglected the boundary effects.
These effects were considered in detail in \cite{L1},\cite{L2},
they can give quite essential contribution in the soft part of
spectrum depending on the target thickness. We neglected also by
effects of the polarization of a medium. They were considered in
detail in \cite{L1}. The relative contribution of polarization of
a medium into probability of pair creation is discussed in
\cite{BK}
\begin{equation}
\frac{\omega_0^2 \varepsilon \varepsilon'}{\omega^2 m^2}
\leq \frac{\omega_0^2}{m^2} < 10^{-7} \ll 1,
\quad \omega_0^2=\frac{4\pi e^2 n_e}{m},
\label{4.5}
\end{equation}
where $n_e$ is the number density of electron in the medium, $\omega_0$
is the plasma frequency. The contribution of polarization of a medium into
the total energy losses in thick target is of the order $\omega_0/m$.
The polarization of a medium affects at the soft part of the spectrum only
at $\omega \leq \omega_p=\gamma \omega_0~(x \leq \omega_p/\varepsilon
=\omega_0/m)$. Even for heavy elements $\omega_0/m \sim 2 \cdot 10^{-4}$.
This contribution was analyzed in \cite{L1}.

\section*{Acknowledgments}. This work was supported in part
by the Russian Fund of Basic
Research under Grant 00-02-18007.

\end{document}